\documentclass[graybox]{svmult}

\usepackage{mathptmx}       
\usepackage{helvet}         
\usepackage{courier}        
\usepackage{type1cm}        
\usepackage{makeidx}         
\usepackage{graphicx}        
\usepackage{multicol}        
\usepackage[bottom]{footmisc}
\usepackage{amsmath,amssymb}
\usepackage{url}
\usepackage{amscd}

\long\def\citat#1{{\narrower\narrower\small  #1 \par \noindent}}
\def\noi{\noindent}

\makeindex             

\begin{document}


\title*{Steps towards {\em Quantum Gravity} and the practice of science: will the merger of
mathematics and physics work?}
\titlerunning{Will the merger of mathematics and physics work?}
\author{Bernhelm Booss-Bavnbek}


\institute{B. Boo{\ss}-Bavnbek \at NSM/IMFUFA, Roskilde Universitet,
Postboks 260, DK-4000 Roskilde, Denmark, \email{booss@ruc.dk},
\url{http://milne.ruc.dk/~Booss/} 
}

\maketitle

\abstract{I recall general tendencies of the mathematization of the
sciences and derive challenges and tentative obstructions for a
successful merger of mathematics and physics on fancied steps
towards {\em Quantum Gravity}. This is an edited version of the
opening words to an international workshop {\em Quantum Gravity: An
Assessment}, Holb{\ae}k, Denmark, May 17-18, 2008. It followed
immediately after the Quantum Gravity Summer School, see
\url{http://QuantumGravity.ruc.dk} }

\section{Regarding the need and the chances of unification}

Reading the literature, preparing and attending the Quantum Gravity
Summer School, editing of this volume, and discussing with
co-editors and contributors provoked the following claims and
questions: (1) The natural laboratory for quantum gravity is the
universe, of which we have no control. (2) Nevertheless, we have to
accept the challenge, and also a new feature: soon a lifetime will
be insufficient to verify a new important theory, and only our
successors will be able to prove or disprove it. By the way, it took
49 years just for the Casimir effect, let alone Hawking radiation or
string theory. (3) Do we really need to unify gravity with the other
interactions? Some physicists say ``No". However, if gravity stays
in isolation, physicists must be afraid that gravitational physics
as a subject will die soon. (4) Most innovative work in quantum
gravity is balancing on a knife's edge between abstract mathematics
and fresh views on physics concepts. A mathematician, however, may
have many remaining unanswered questions, both regarding the claims
of physics relevance of the most innovative work and regarding the
mathematical clearness and reliability of various new concepts and
calculations.

In the following I recall common knowledge on modelling,
mathematization, and science history to put the declared ``New Paths
Towards Quantum Gravity" in a common frame in spite of their
scattering and heterogeneity. Some of the following considerations
were published in \cite{BooEspLes:QG} in condensed form before the
Summer School as a kind of {\em platform} for the assessment of our
endeavour. For a comparison with mathematization in other frontier
fields I refer also to the recent \cite{Boo:MIS}.

\section{The place of physics in John Dee's {\em Groundplat of
Sciences and Artes, Mathematicall} of 1570}

The use of mathematical arguments, first in pre-scientific
investigations, then in other sciences, foremost in medicine and
astronomy and in their shared border region astrology, has been
traced way back in history by many authors from various
perspectives, Bernal \cite{Ber:SIH}, H{\o}yrup \cite{Hoy:FHS}, and
Kline \cite{Kli:MTF}.

Globally speaking, they all agree on three mathematization
tendencies:

\begin{enumerate}

\item The progress in the individual sciences makes work on ever
more complicated problems possible and necessary.

\item This accumulation of problems and data demands conscious,
planned, and economic procedures in the individual sciences, i.e.,
an increased emphasis on questions of methodology.

\item This increased emphasis on questions of methodology is as a
rule associated with the tendency of mathematization.

\end{enumerate}

All of this applies generally. In detail, we find many various
pictures. In his {\em Groundplat of Sciences and Artes,
Mathematicall} of 1570 \cite{Dee:GSA}, the English alchemist,
astrologer, and mathematician John Dee, the first man to defend the
Copernican theory in Britain and a consultant on navigation, pointed
out, in best Aristotelian tradition, that it is necessary in the
evaluation of mathematization to pay strict attention to the
specific characteristics of the application area in question. He
postulated a dichotomy between the {\em Principall} side, pure
mathematics, and the {\em Deriuative} side, i.e., applied
mathematics and mathematization. He then classified the applications
of pure mathematics according to objects treated:

\begin{itemize}

\item {\em Ascending Application in thinges Supernaturall, eternall and Diuine},

\item {\em In thinges Mathematicall: Without farther Applications},

\end{itemize}
and finally, on the lowest and most vulgar plane in the Aristotelian
scheme,

\begin{itemize}

\item {\em Descending Application in thinges Naturall: both
Substantiall \& Accidentall, Visible \& Inuisible \& c.}.

\end{itemize}
Now that history has excluded matters divine from mathematics, we
can with some justification ask whether later generations may regard
with equal amusement and astonishment the fact that in our time
there are a large number of professional mathematicians and
physicists, who are completely satisfied with spending their entire
lives working in the second, inner mathematical level and who
persistently refuse to descend to vulgar applications.

The panorama of the individual sciences and the role that
mathematics had to play in them was perfectly clear to John Dee. In
our time the matter is somewhat more complex. For a class on Quantum
Gravity, I cannot point out a geodetically perfect picture of
today's landscape of mathematization nor of precise border lines
between the mathematics and physics addressed. I must treat the
matter rather summarily. A summary treatment may have the advantage
that in comparison among the mathematization progress in different
branches of physics, common problems on one hand and on the other
hand special features of the here advocated new paths towards
quantum gravity can be seen more clearly.

In the following, I shall restrict myself to the study of dead
nature in {\em physics}, the field which has the highest degree of
mathematization on any chosen scale, both quantitatively and
qualitatively. To put things in relief, I shall occasionally touch
upon the investigation of living matter in {\em medicine}, the field
where one might expect the greatest mathematization advances in our
century, and confront our highly speculative branch of mathematical
physics with the treatment of financial issues and decision making
for commerce and production in {\em economics}, a field of
questionable scientific state, that, beyond well founded actuary
estimations, lacks unambiguous results and convincing clear
perspectives regarding mathematization.


\section{Delimitation between mathematics and physics}\label{s:physics}

The intimate connection between mathematics and physics makes it
difficult to determine the theoretical relevance of mathematics and
obscures the boundary between genuinely physical thought and
observation on one side and the characteristically mathematical
contribution on the other side. Recall Hilbert's perception of
probability theory as a chapter of physics in his famous 6th Problem
\cite{Hil:MP}:

\citat{6. Mathematical Treatment of the Axioms of Physics. The
investigations on the foundations of geometry suggest the problem:
{\em To treat in the same manner, by means of axioms, those physical
sciences in which already today mathematics plays an important part;
in the first rank are the theory of probabilities and mechanics}.}

\noi To say it mildly, as Gnedenko did in his comments to the
Russian edition of 1969: {\em Today this viewpoint} (to consider
probability theory as a chapter of physics) {\em is no longer so
common as it was around the turn of the century, since the
independent mathematical content of the theory of probabilities has
sufficiently clearly showed since then...} With hindsight and in
view of the still challenging foundational problems of quantum
mechanics, however, we may accept that parts of mathematics and
physics can be interlaced in a non-separable way.

Another famous example of that inextricable interlacement is
provided by the Peierls-Frisch memorandum of 1940 to the British
Government: suggested by the codiscoverer of fission Otto Frisch,
the physicist Rudolph Peierls, like Frisch a refugee in Britain,
made the decisive feasibility calculation that not tons (as -
happily - erroneously estimated by Heisenberg in the service of the
Nazis) but only about 1 kg (later corrected to 6 kg) of the pure
fissile isotope $U_{235}$ would be needed to make the atomic bomb.
Was it mathematics or physics? It may be worth mentioning that
Peierls was a full professor at the University of Birmingham since
1937 and became joint head of {\em mathematics} there,
\cite{Edw:REP}. Theoretical physics in Britain is often in
mathematics. As a matter of fact, physics in our sense did not exist
as a single science before the nineteenth century. There were
well-defined {\em experimental physics} comprising heat, magnetism,
electricity and colour, leaving mechanics in mathematics, see
\cite[p. 493]{Hoy:FHS}.

 In spite of that intermingling, physics can provide a ready system
 of categories to distinguish different use of mathematics in
 different modelling situations. Perhaps, the situation can be best
 compared with the role of physics in general education. After all,
 physics appears as the model of mathematization: there is no
 physics without mathematics - and, as a matter of fact, learning of
 mathematics is most easy in a physics context: calculation by
 letters; the various concepts of a function (table, graph,
 operation) and its derivatives and anti-derivatives; differential
 equations; the concept of observational errors and the
 corresponding estimations and tests of hypotheses; Brownian
 movements; all these concepts can be explained context-free or in
 other contexts (where some of the concepts actually originated),
 but they become clearest in the ideally simple applications of
 physics, which are sufficiently complicated to see the superiority
 of the mathematization as compared to feelings, qualitative
 arguments, discussions, convictions, imagination - but simple
 enough to get through.

\section{Variety of modelling purposes}\label{ss:variety}

It may be helpful to distinguish the following modelling purposes:

\subsection{Production of data, model-based measurements} Clearly,
the public associates the value of mathematical modelling foremost
to its {\em predictive} power, e.g., in numerical weather
prediction, and its {\em prescriptive} power, e.g., in the design of
the internal ballistics of the hydrogen bomb; more flattering to
mathematicians, the {\em explanatory} power of mathematization and
its contribution to {\em theory development} yield the highest
reputation within the field. However, to the progress of physics,
the {\em descriptive} role, i.e., supporting model-based
measurements in the laboratory, is -- as hitherto -- the most
decisive contribution of mathematics. Visco-elastic constants and
phase transition processes of glasses and other soft materials can
not be measured directly. For high precision in the critical region,
one measures electric currents through a ``dancing" piezoelectric
disc with fixed potential and varying frequency. In this case,
solving mathematical equations from the fields of electro-dynamics
and thermo-elasticity becomes mandatory for the design of the
experiments and the interpretation of the data. In popular terms,
one may speak of a {\em mathematical microscope}, in technical terms
of a {\em transducer} that becomes useful as soon as we understand
the underlying mathematical equations.

\subsection{Simulation}\label{ss:simulation} Once a model is found
and verified and the system's parameters are estimated for one
domain, one has the hope of doing computer calculations to predict
what new experiments in new domains (new materials, new temperatures
etc) should be made  and what they might be expected to show.
Rightly, one has given a special name of honour to that type of
calculations, {\em computer simulations}: as a rule, it requires to
run the process on a computer or a network of computers under quite
sophisticated conditions: typically, the problem is to bring the
small distances and time intervals of well-understood molecular
dynamics up to reasonable macroscopic scales, either by aggregation
or by Monte Carlo methods -- as demonstrated by Buffon's needle
casting for the numerical approximation of $\pi$.

One should be aware that the word {\em simulation} has, for good and
bad, a connotation derived from NASA's space simulators and
Nintendo's war games and jukeboxes. Animations and other advanced
computer simulations can display an impressive beauty and convincing
power. That beauty, however, is often their dark side: simulations
can show a deceptive similarity with true observations, so in
computational fluid dynamics when the numerical solution of the
Bernoulli equations, i.e., the linearization of the Navier-Stokes
equations for laminar flow displays eddies characteristic for the
non-linear flow. The eddies do not originate from real energy loss
due to friction and viscosity but from hardly controllable hardware
and software properties, the chopping of digits, thus providing a
{\em magic realism}, as coined by Abbott and Larsen
\cite{AbbLar:MCD}. In numerical simulation, like in mathematical
statistics, results which fit our expectations too nicely, must
awake our vigilance instead of being taken as confirmation.


\subsection{Prediction}\label{ss:prediction} As shown in the
preceding subsection, there is no sharp boundary between description
and prediction. However, the quality criteria for predictions are
quite simple: do things develop and show up as predicted? So, for
high precision astrology and longitudinal determination in deep-sea
shipping the astronomical tables of planetary movement, based on the
outdated and falsified Ptolemaic system (the {\em Resolved Alfonsine
Tables}) and only modestly corrected in the {\em Prutenic Tables} of
1551 were, until the middle of the 17th century, rightly considered
as more reliable than Kepler's heliocentric {\em Rudolphine Tables},
as long as they were more precise - no matter on what basis, see,
Steele \cite[p. 128]{Ste:OPE}.

Almost unnoticed, we have had a similar revolution in weather
prediction in recent years: the (i) analogy ({\it synoptic}) methods
of identifying a similarly looking weather situation in the weather
card archives to base the extrapolation on it were replaced by
almost pure (ii) numerical methods to derive the prediction solely
from the thermodynamic and hydrodynamic basic equations and
conservation laws, applied to initial conditions extracted from the
observation grid. ``Almost" because the uncertainty of the
interpolation of the grid and the high sensitivity of the evolution
equations to initial conditions obliges to repeated runs with small
perturbations and human inspection and selection of the most
``probable" outcome like in (i). That yields sharp estimates about
the certainty of the prediction for a range of up to 10 days. In
nine of ten cases, the predictions are surprisingly reliable and
would have been impossible to obtain by traditional methods.
However, a 10{\%} failure rate would be considered unacceptable in
industrial quality control.

In elementary particle physics, the coincidence of predictions with
measurements is impressive, but also disturbing. I quote from Smolin
\cite[pp. 12-13]{Smo:TWP}:

\citat{Twelve particles and four forces are all we need to explain
everything in the known world. We also understand very well the
basic physics of these particles and forces. This understanding is
expressed in terms of a theory that accounts for all these particles
and all of the forces except for gravity. It's called {\em the
standard model of elementary-particle physics} - or the standard
model for short. \dots Anything we want to compute in this theory we
can, and it results in a finite number. In the more than thirty
years since it was formulated, many predictions made by this theory
have been checked experimentally. In each and every case, the theory
has been confirmed.

The standard model was formulated in the early 1970s. Except for the
discovery that neutrinos have mass, it has not required adjustment
since. So why wasn't physics over by 1975? What remained to be done?

For all its usefulness, the standard model has a big problem: It has
a long list of adjustable constants. \dots}

\noi We feel pushed back to the pre-Keplerian, pre-Galilean and
pre-Newtonian cosmology built on ad-hoc assumptions, displaying
clever and deceptive mathematics-based similarity \index{deceptive
similarity} between observations and calculations -- and ready to
fall at any time because the basic assumptions are not explained.

Perhaps the word {\em deceptive} is inappropriate when speaking of
description, simulation and prediction: for these tasks, {\em
similarity} can rightly be considered as the highest value
obtainable, as long as one stays in a basically familiar context.
From a semiotic angle, the very similarity must have a meaning and
is indicating something; from a practical angle, questions regarding
the epistemological status can often be discarded as metaphysical
exaggerations: who cares about the theoretical or ad-hoc basis of a
time schedule in public transportation -- as long as the train
leaves on time!

 \subsection{Control}\label{ss:control} The prescriptive power of
 mathematization deserves a more critical examination. In physics
 and engineering we may distinguish between the (a) feasibility, the
 (b) efficiency, and the (c) safety of a design. A {\em design} can
 be an object like an airplane or a circuit diagram for a chip, an
 instrument like a digital thermometer, TV set, GPS receiver or
 pacemaker, or a regulated process like a feed-back regulation of
 the heat in a building, the control of a power station or the
 precise steering of a radiation canon in breast cancer therapy.
 Mathematics has its firm footing for testing (a) in thought
 experiments, estimations of process parameters, simulations and
 solving equations. For testing (b), a huge inventory is available
 of mathematical quality control and optimization procedures by
 variation of key parameters.

It seems to me, however, that (c), i.e., safety questions provide
the greatest mathematical challenges. They appear differently in (i)
experience-based, (ii) science-based and (iii) science-integrated
design. In (i), mathematics enters mostly in the certification of
the correctness of the design copy and the quality test of the
performance. In (ii), well-established models and procedures have to
be modified and re-calculated for a specific application.
Experienced physicists and engineers, however, seldom trust their
calculations and adaptations. Too many parameters may be unknown and
pop-up later: Therefore, in traditional railroad construction, a
small bridge was easily calculated and built, but then
photogrammetrically checked when removing the support constructions.
A clash of more than $\delta_{\operatorname{crit}}$ required
re-building. Similarly, even the most carefully calculated chemical
reactors and other containers under pressure and heat have their
prescribed ``Soll-Bruchstelle" (supposed line of fracture) in case
that something is going wrong.

The transition from (ii) to (iii) is the most challenging: very
seldom one introduces a radically new design in the physics
laboratory or engineering endeavour. But there are systems where all
components and functions can be tested separately though the system
as a whole can only be tested {\em in situ}: a new design of a
Diesel ship engine; a car, air plane or space craft; a new concept
in cryptography. In all these cases, one is tempted to look and even
to advocate for mathematical proofs of the safe function according
to specification. Unfortunately, in most cases these ``proofs"
belong rather to the field of fiction than to rigorous mathematics.
For an interesting discussion on ``proofs" in cryptography (a little
remote from physics) see the debate between Koblitz and opponents in
\cite{Kob:xxx} and follow-ups in the {\em Notices of the American
Mathematical Society}.

An additional disturbing aspect of science-integrated technology
development is the danger of a loss of transparency. Personally, I
must admit, I am grateful for most black-box systems. I have no
reason to complain when something in my computer is hidden for my
eyes, as long as everything functions as it shall or can easily be
re-tuned. However, for the neighbourhood of a chemical plant (and
the reputation of the company) it may be better not to automatize
everything but to keep some aspects of the control non-mathematized
and in the hands of the service crew to avoid de-qualification and
to keep the crew able to handle non-predictable situations.

A last important aspect of the prescriptive power of the
mathematization is its formatting power for thought structure and
social behaviour. It seems that there is not so much to do about it
besides being aware of the effects.

 \subsection{Explain phenomena} The noblest role of mathematical
 concepts in physics is to explain phenomena. Einstein did it when
 {\em reducing} the heat conduction to molecular diffusion, starting
 from the formal analogy of Fick's Law with the cross section of
 Brownian motion. He did it also when {\em generalizing} the
 Newtonian mechanics into the special relativity of constant light
 velocity and again when {\em unifying} forces and curvature in
 general relativity.

Roughly speaking, mathematical models can serve physics by reducing
new phenomena to established principles; as heuristic devices for
suitable generalizations and extensions; and as ``a conceptual
scheme in which the insights \dots fit together" (C. Rovelli).
Further below I shall return to the last aspect -- the unification
hope.

Physics history has not always attributed the best credentials to
explaining phenomena by abstract constructions. It has discarded the
concept of a ghost for perfect explanation of midnight noise in old
castles; the concept of ether for explaining the finite light
velocity; the phlogiston for burning and reduction processes, the
Ptolemaic epicycles \index{Ptolemaic system} for planetary motion.
It will be interesting to see in the years to come whether the
mathematically advanced String Theory or the recent Connes-Marcolli
reformulation of the Standard Model in terms of spectral triples
will undergo the same fate.

 \subsection{Theory development} Finally, what has been the role of
 mathematical concepts and mathematical beauty for the very theory
 development in physics? One example is Johann Bernoulli's purely
 aesthetic confirmation of Galilean fall law $s\,=\, g/2 \ t^2$
 among a couple of candidates as being the only one providing the
 same equation (shape) for his brachistochrone and Huygens'
 tautochrone, \cite[p. 395]{Ber:xxx}:

\citat{Before I end I must voice once more the admiration that I
feel for the unexpected identity of Huygens' tautochrone and my
brachistochrone. I consider it especially remarkable that this
coincidence can take place only under the hypothesis of Galilei, so
that we even obtain from this a proof of its correctness. Nature
always tends to act in the simplest way, and so it here lets one
curve serve two different functions, while under any other
hypothesis we should need two curves.}

Another, more prominent example is the lasting triumph of Maxwell's
equations: a world of radically new applications was streaming out
of the beauty and simplicity of the equations of electro-magnetic
waves!

However, not every mathematical, theoretical and empirical
accumulation leads to theory development. Immediately after
discovering the high-speed rotation of the Earth around its own
axis, a spindle shape of the Earth was suggested and an
infinitesimal tapering towards the North pole confirmed in geodetic
measurements around Paris. Afterwards, careful control measurements
of the gravitation at the North Cap and at the Equator suggested the
opposite, namely an ellipsoid shape with flattened poles. Ingenious
mathematical mechanics provided a rigorous reason for that. Gauss
and his collaborator Listing, however, found something different in
their control. They called the shape {\em gleichsam
wellenf{\"o}rmig} and dropped the idea of a theoretically
satisfactory description. Since then we speak of a {\em Geoid}. For
details see Listing \cite{Lis:KGG} and the receent Torge \cite[p.
3]{Tor:G}.

 \section{``The trouble with physics"}

That is the title of an interesting and well-informed polemic by Lee
Smolin against String Theory and present main stream physics at
large. He notices a {\em stagnation} in physics, {\em so much
promise, so little fulfillment} \cite[p. 313]{Smo:TWP}, a
predominance of anti-foundational spirit and contempt for visions,
partly related to the mathematization paradigm of the 1970s,
according to Smolin:  {\em Shut up and calculate}.

Basically, Smolin may be right. B{\o}rge Jessen, the Copenhagen
mathematician and close collaborator of Harald Bohr once suggested
to distinguish in sciences and mathematics between periods of {\em
expansion} and periods of {\em consolidation}. Clearly physics had a
consolidation period in the first half of the 20th century with
relativity and quantum mechanics. The same may be true for biology
with the momentous triumph of the DNA disclosure around 1950, while,
to me, the mathematics of that period is characterized by an almost
chaotic expansion in thousands of directions. Following that way of
looking, mathematics of the second half of the 20th century is
characterized by an enormous consolidation, combining so disparate
fields like partial differential equations and topology in index
theory, integral geometry and probability in point processes, number
theory, statistical mechanics and cryptography, etc. etc. A true
period of consolidation for mathematics, while - at least from the
outside - one can have the impression that physics and biology of
the second half of the 20th century were characterized merely by
expansion, new measurements, new effects - and almost total absence
of consolidation or, at least failures and vanity of all trials in
that direction.

Indeed, there have been impressive successes in recent physics, in
spite of the absence of substantial theoretical progress in physics:
perhaps the most spectacular and for applications most important
discovery has been the High Temperature Superconducting (HTS)
property of various ceramic materials by Bednorz and M{\"u}ller -
seemingly without mathematical or theoretical efforts but only by
systematic combinatorial variation of experiments - in the tradition
of the old alchemists, \cite{BedMul:HTS}.

The remarkable advances in fluid dynamics, weather prediction,
oceanography, climatic modelling are mainly related to new
observations and advances in computer power while the equations have
been studied long before.

Nevertheless, I noticed a turn to theory among young experimental
physicists in recent years, partly related to investigating the {\em
energy landscapes} in material sciences, partly to the re-discovery
of the {\em interpretational} difficulties of quantum mechanics in
recent quantum optics.

 \section{Theory -- model -- experiment}

Physics offers an extremely useful practical distinction between
{\em theory}, {\em model} and {\em experiments}. From his deep
insight in astronomy, computing, linguistics and psychology, Peter
Naur ridicules such distinctions as ``metaphysical exaggeration" in
\cite{Nau:KML}. He may be right. We certainly should not exaggerate
the distinction. In this review, however, the distinction helps to
focus on differences of the role of mathematics in doing science.

\subsection{First principles} By definition, the very core of
modelling is mathematics. Moreover, if alone by the stochastic
character of observations, but also due to the need to understand
the mathematics of all transducers involved in measurements,
mathematics has its firm stand with experiments. First principles,
however, have a different status: they do not earn their authority
from the elegance of being mathematically wrapped, but from the
almost infinite repetition of similar and, as well disparate
observations connected to the same principle(s). In the first
principles, mathematics and physics meet almost on eye level: first
principles are also {\em established} - like mathematics, and are
only marginally questioned. To me, the problem with the pretended
eternal authority of first principles is that new cosmological work
indicates that the laws of nature may also have undergone some
development; that there might have ``survived" some evolutionary
relics; and that we had better be prepared to be confronted under
extreme experimental conditions, with phenomena and relations which
fall out of the range of accredited first principles. The canonical
candidate for such a relic is the Higgs particle, whether already
observed or not. Participants of the Quantum Gravity Assessment
Workshop 2008 will recall Holger Bech Nielsen's contributions.

\subsection{Towards a taxonomy of models}Not necessarily for the
credibility of mathematical models, but for the way of checking the
range of credibility, the following taxonomy of models may be
extremely useful.

The Closing Round Table of the International Congress of
Mathematicians (Mad\-rid, August 22-29, 2006) was devoted to the
topic {\em Are pure and applied mathematics drifting apart?} As a
panelist, Yuri Manin subdivided the mathematization, i.e., the way
mathematics can tell us something about the external world, into
three modes of functioning (similarly Bohle, Boo{\ss} and Jensen
1983, \cite{BoBoJe:IEO}, see also \cite{Bo:AIF}):

\begin{enumerate}

\item An {\em (ad-hoc, empirically based) mathematical model}
\index{ad-hoc model} ``describes a certain range of phenomena,
qualitatively or quantitatively, but feels uneasy pretending to be
something more." Manin gives two examples of the predictive power of
such models, Ptolemy's model of epicycles \index{Ptolemaic system}
describing planetary motions of about 150 BCE, and the standard
model of around 1960 describing the interaction of elementary
particles, besides legions of ad-hoc models which hide the lack of
understanding behind a more or less elaborated mathematical
formalism of organizing available data.

\item A {\em mathematically formulated theory} is distinguished from
an ad-hoc model primarily by its ``higher aspirations. A theory, so
to speak, is an aristocratic model." Theoretically substantiated
models, such as Newton's mechanics, are not necessarily more precise
than ad-hoc models; the coding of experience in the form of a
theory, however, allows a more flexible use of the model, since its
embedding in a theory universe permits a theoretical check of at
least some of its assumptions. A theoretical assessment of the
precision and of possible deviations of the model can be based on
the underlying theory.

\item A {\em mathematical metaphor} postulates that ``some complex
range of phenomena might be compared to a mathematical
construction". As an example, Manin mentions artificial intelligence
with its ``very complex systems which are processing information
because we have constructed them, and we are trying to compare them
with the human brain, which we do not understand very well -- we do
not understand almost at all. So at the moment it is a very
interesting mathematical metaphor, and what it allows us to do
mostly is to sort of cut out our wrong assumptions. If we start
comparing them with some very well-known reality, it turns out that
they would not work."

\end{enumerate}

Clearly, Manin noted the deceptive formal similarity of the three
ways of mathematization which are radically different with respect
to their empirical foundation and scientific status. He expressed
concern about the lack of distinction and how that may ``influence
our value systems". In the words of \cite[p. 73]{Bo:AIF}:

\citat{Well founded applied mathematics generates prestige which is
inappropriately generalized to support these quite different
applications. The clarity and precision of the mathematical
derivations here are in sharp contrast to the uncertainty of the
underlying relations assumed. In fact, similarity of the
mathematical formalism involved tends to mask the differences in the
scientific extra-mathematical status, in the credibility of the
conclusions and in appropriate ways of checking assumptions and
results... Mathematization can -- and therein lays its success --
make existing rationality transparent; mathematization cannot
introduce rationality to a system where it is absent... or
compensate for a deficit of knowledge.}

\noi Asked whether the last 30 years of mathematics' consolidation
raise the chance of consolidation also in phenomenologically and
metaphorically expanding sciences, Manin hesitated to use such
simplistic terms. He recalled the notion of Kolmogorov complexity of
a piece of information, which is, roughly speaking,

\citat{the length of the shortest programme, which can be then used
to generate this piece of information... Classical laws of physics
-- such phantastic laws as Newton's law of gravity and Einstein's
equations -- are extremely short programmes to generate a lot of
descriptions of real physical world situations. I am not at all sure
that Kolmogorov's complexity of data that were uncovered by, say,
genetics in the human genome project, or even modern cosmology data
... is sufficiently small that they can be really grasped by the
human mind.}

\subsection{The scientific status of quantum gravity as compared to medicine and economics}
From the rich ancient literature preserved, see Diepgen
\cite{Die:GME}, Kudlien \cite{Kud:BMD} and, in particular, J{\"u}rss
\cite[312--315]{Jur:GWD}, we can see that the mind set in Greek
medicine already from the fifth century BCE was {\em ours}: instead
of the partition (familiar from earlier and shaman medicine and
similar to the mind set preserved, as seen above, in physics until
recent times) into an empirical-rational branch (healing wounds) and
a religious-magic branch (cure inner diseases), a physiological
concept emerged which focused on the patient as an individual
organism within a population, with organs, liquids and tissue,
subjected to environmental and dietetic influences and, in
principle, open for unconfined investigation of functions, causal
relations and the progressive course of diseases. In Hippocratic
medicine, we meet for the first time the visible endeavour after a
rational surmounting of all problems related to body events.

With a shake of the head, we may read of Greek emphasis and
speculations about the body's four liquids or other strange things,
like when we recall today the verdict of the medical profession 60
years ago against drinking water after doing sports and under
diarrhoea or their blind trust in antibiotics, not considering
resistance aspects at all. Admittedly, we have no continuity of
results in medicine, but, contrary to physics, we have an outspoken
continuity in mind set: no ghosts, no metaphysical spirits, no
fancied particles or relations are permitted to enter our
explanations, diagnoses, prevention, cure and palliation.

Physicists of our time like to date the physics' beginning back to
Galileo Galilei and his translation of measurable times and
´distances on a skew plane into an abstract fall law. Before Galilei
- and long time after him, the methodological scientific status of
what we would call mechanical physics was quite low as compared with
medicine. Physics was a purely empirical subject. It was about
precise series of observations and quantitative extrapolations. It
was the way to predict planetary positions, in particular eclipse
times, the content of silver in compounds, or the manpower required
to lift a given weight with given weight arm. It was accompanied and
mixed up with all kinds of speculations about the spirits and ghosts
at work. We can easily see the continuity of results, of
observations and calculations from Kepler and Newton to our time.
However, we can hardly recognize anything in their thinking about
physics, in the way they connected physics with cosmic music or
alchemy or formulated assumptions. We may wonder what later
generations will think about our fancied new paths towards quantum
gravity.

While a rational point of departure for economics, in particular
under the present crisis, can only be a {\em systems view}, a
holistic unifying view in physics like our efforts in quantum
gravity have a smell of vanity. One may argue that the time has
hardly come for that endeavour - comparable to the felt necessity
but still continuing futility of or at least doubts about a holistic
all-embracing systems biology programme in medicine.

\section{General trends of mathematization and modelling}

\subsection{Deep divide}\label{ss:dd}
Regarding the power and the value of mathematization, there is a
deep moral divide both within the mathematics community and the
public.

On the one side, we have the outspoken science and math optimism of
outstanding thinkers: Henri Poincar{\'e}'s {\em Nature not only
suggests to us problems, she suggests their solution}; David
Hilbert's {\em Wir m{\"u}ssen wissen; wir werden wissen - We must
know; we will know} of his Speech in K{\"o}nigsberg in 1930, now on
his tomb in G{\"o}ttingen; or Bertolt Brecht's vision of
mathematical accountability in {\em Die Tage der Kommune}
\cite{Bre:DTK} of 1945: ``Das ist die Kommune, das ist die
Wissenschaft, das neue Jahrtausend... - That is the Commune, that is
the science, the new millennium..."). We have astonishing evidence
that many mathematization concepts either appear to us as {\em
natural} and {\em a-priori}, or they use to emerge as clear over
time. We have the power and validity of extremely simple concepts,
as in {\em dimension analysis}, {\em consistency requirements} and
{\em gauge invariance} of mathematical physics. Progressive
movements emphasize science and education in liberation movements
and developing countries. Humanitarian organizations (like WHO and
UNICEF) preach science and technology optimism in confronting mass
poverty and epidemics.

On the other side, deep limitation layers of science and
mathematical thinking have been dogged up by Kurt G{\"o}del's {\em
Incompleteness Theorem} for sufficiently rich arithmetic systems,
Andrei N. Kolmogorov's {\em Complexity Theory}, and Niels Bohr's
notion of {\em Complementarity}. Incomprehensibility and lack of
regularity continue to hamper trustworthy mathematization. Peter Lax
\cite[p. 142]{Lax:MAP} writes about the {\em profound mystery of
fluids}, though recognizing that different approaches lead to
remarkable coincidence results, supporting reliability.

The abstruseness of the mathematical triumphs of the hydrogen bomb
is commonplace. The wide-spread trust in superiority and
invincibility, based on mathematical war technology like high
precision bombing, has proved to be even more vicious for warriors
and victims than the immediate physical impact of the very
math-based weaponry, recently also in Iraq and Afghanistan.

In between the two extremes, Hilbert's optimistic prediction of {\em
clearness} and the sceptical Kafkaesque expectation of increasing
{\em bewilderment} when digging deeper mathematically, we have the
optimistic scepticism of Eugene Wigner's {\em unreasonable
effectiveness of mathematics}, but also Jacob Schwartz's verdict
against the {\em pernicious influence of mathematics on science} and
Albert Einstein's demand for {\em finding the central questions}
against the dominance of the beautiful and the difficult.



\subsection{Charles Sanders Peirce's semiotic view}
From the times of Niels Bohr, many physicists, mathematicians and
biologists have been attentive to philosophical aspects of our
doing. Most of us are convinced that the frontier situation of our
research can point to aspects of some philosophical relevance - if
only the professional philosophers would take the necessary time to
become familiar with our thinking. Seldom, however, we read
something of the philosophers which can inspire us.

The US-American philosopher Charles Sanders Peirce (1839-1914) is an
admirable exception. In his semiotics and pragmaticist (he avoids
the word ``pragmatic") thinking, he provides a wealth of ideas,
spread over an immense life work. It seems to me that many of his
ideas, comments, and concepts can shed light on the why and how of
mathematization. Here I shall only refer some thoughts of Peirce's
{\em The Fixation of Belief} from 1877, see \cite{Pei:FB}.

My fascination of Peirce's text is, in particular, based on the
following observations which may appear trivial (or known from
Friedrich Engels), but are necessary to repeat many times for the
new-modeller:


\noi 1. For good and bad, we are all equipped with innate (or
spontaneous) orientation, sometimes to exploit, sometimes to subdue.
Our innate orientation is similar to the  habits of animals in our
familiar neighbourhood. We are all "logical machines".


\noi 2. However, inborn logic is not sufficient in foreign (new)
situations. For such situations, we need methods how to fixate our
beliefs. Peirce distinguishes four different methods. All four have
mathematical aspects and are common in mathematical modelling.
\begin{description}

\item[Tenacity] is our strength not to become confused, not to be
blown away by unfounded arguments, superficial objections,
misleading examples, though sometimes keeping our ears locked for
too long.

\item[Authority] of well-established theories and results is what we tend to believe in
and have to stick to. We will seldom drop a mastered approach in
favour of something new and unproved.

\item[Discussion] can hardly help to overcome a belief built on tenacity or authority.

\item[Consequences] have to be investigated in all modelling. At the end of the day,
they decide whether we become convinced of the validity of our
approach (Peirce's Pragmaticist Maxim).

\end{description}


\noi 3. The main tool of modelling (i.e., the fixation of belief by
mathematical arguments) is the transformation of symbols (signals,
observations, {\em segments of reality}) into a new set of symbols
(mathematical equations, {\em models} and {\em descriptions}). The
advantage for the modeller, for the person to interpret the signs,
is that signs which are hard or humid and difficult to collect in
one hand can be replaced by signs which we can write and manipulate.


\noi 4. The common mapping cycle {\em reality} $\to$ {\em model}
$\to$ {\em validation} is misleading. The quality of a mathematical
model is not how similar it is to the segment of reality under
consideration, but whether it provides a flexible and goal-oriented
approach, opening for doubts and indicating ways for the removal of
doubts (later trivialized by Popper's falsification claim). More
precisely, Peirce claims

\begin{itemize}

\item
Be aware of differences between different approaches!

\item
Try to distinguish different goals (different priorities) of
modelling as precise as possible!

\item
Investigate whether different goals are mutually compatible, i.e.,
can be reached simultaneously!

\item

Behave realistically! Don't ask: {\em How well does the model
reflect a given segment of the world}? But ask: {\em Does this model
of a given segment of the world support the wanted and possibly
wider activities / goals better than other models}?
\end{itemize}

I may add: we have to strike a balance between \textit{Abstraction
vs. construction}, \textit{Top-down vs. bottom-up}, and
\textit{Unification vs. specificity}. We better keep aware of the
variety of \textit{Modelling purposes} and the multifaceted
relations between \textit{Theory - model - experiment}. Our
admiration for the \textit{Power of mathematization}, the
\textit{Unreasonable effectiveness of mathematics} (Wigner) should
not blind us for the \textit{Staying and deepening limitations of
mathematization opposite new tasks}.



\end{document}